\documentclass[preprint,showpacs,preprintnumbers,amsmath,amssymb]{revtex4}

\usepackage{graphicx}
\usepackage{dcolumn}
\usepackage{bm}

\providecommand{\Journal}[4] {#1 {\bf #2}, #3 (#4)}
\providecommand{\EPJC}{Eur. Phys. J. C } %
\providecommand{\LNC}{Lett. Nuovo Cimento } %
\providecommand{\MPLA}{Mod. Phys. Lett. A} %
\providecommand{\PLB}{Phys. Lett. B } %
\providecommand{\PRL}{Phys. Rev. Lett. } %
\providecommand{\PRD}{Phys. Rev. D } %
\providecommand{\ZPC}{Z. Phys. C } %

\begin{document}

\title{Energy scale independence of Koide's relation for quark and lepton masses}

\author{Nan Li}
\affiliation{School of Physics, Peking University, Beijing 100871,
China}
\author{Bo-Qiang Ma}\altaffiliation{Corresponding author}\email{mabq@phy.pku.edu.cn}
\affiliation{ CCAST (World Laboratory), P.O.~Box 8730, Beijing
100080, China \\
and School of Physics, Peking University, Beijing 100871,
China\footnote{Mailing address}}

\begin{abstract}
Koide's mass relation of charged leptons has been extended to
quarks and neutrinos, and we prove here that this relation is
independent of energy scale in a huge energy range from
$1~\mbox{GeV}$ to $2\times10^{16}~\mbox{GeV}$. By using the
parameters $k_u$, $k_d$ and $k_{\nu}$ to describe the deviations
of quarks and neutrinos from the exact Koide's relation, we also
check the quark-lepton complementarity of masses such as $
k_{l}+k_{d} \approx k_{\nu}+k_{u} \approx 2$, and show that it is
also independent (or insensitive) of energy scale.
\end{abstract}


\maketitle

\section{Introduction}

Despite splendid successes of the standard model of particle
physics, some fundamental questions remained unanswered, among which
the origin of fermion masses is one of the most important problems.
In the standard model, these masses are taken as free parameters,
and can only be extracted from experimental data. Phenomenological
analysis aiming at discovering the numerical relations between
fermion masses is useful and practical for the exploration of the
mystery of lepton and quark masses. There have been some conjectures
on this problem (for example,~\cite{barut,hdv}), in which Koide's
relation~\cite{koide,koide2} is the most accurate one, which links
the masses of charged leptons together,
\begin{equation}\nonumber
m_e+m_{\mu}+m_{\tau}=\frac{2}{3}(\sqrt{m_e}+\sqrt{m_{\mu}}+\sqrt{m_{\tau}})^2,
\end{equation}
where $m_e$, $m_{\mu}$, $m_{\tau}$ are the masses of electron,
muon, and tau, respectively. In order to see the accuracy of
Koide's relation, we can introduce a parameter $k_l$,
\begin{equation}
k_l\equiv\frac{m_e+m_{\mu}+m_{\tau}}{\frac{2}{3}(\sqrt{m_e}+\sqrt{m_{\mu}}+\sqrt{m_{\tau}})^2}.\label{eq.kl}
\end{equation}
With the data of PDG~\cite{PDG},
$m_e=0.510998902\pm0.000000021~\mbox{MeV}$,
$m_{\mu}=105.658357\pm0.000005~\mbox{MeV}$ and
$m_{\tau}=1776.99^{+0.29}_{-0.26}~\mbox{MeV}$, we can get the range
of $k_l$, $k_l=1^{+0.00002635}_{-0.00002021}$, which is perfectly
close to 1.

This relation was speculated on the basis of a composite model of
quarks and leptons~\cite{koide} and the extended technicolor-like
model~\cite{koide2}. The fermion mass matrix in these models is
taken as
\begin{equation}\nonumber
M_f=m_0^fGO_f G,
\end{equation}
where $G=\mbox{diag}(g_1,g_2,g_3)$. With the assumptions
$g_i=g^{(1)}+g_{i}^{(8)}$, $\sum_{i}g_{i}^{(8)}=0$ and
$\sum_{i}(g_{i}^{(8)})^2=3(g_{i}^{(1)})^2$, and the charged lepton
mass matrix is $3\times3$ unit matrix, Koide obtained this relation.
This relation is deduced in several other different models by Koide
also~\cite{koideother}. (For a review , see~\cite{review}.) Several
explanations for this excellent relation were presented in the past
ten years. Foot~\cite{foot} gave a geometrical interpretation to it,
\begin{equation}\nonumber
\cos\theta_l=\frac{(\sqrt{m_e} , \sqrt{m_{\mu}} , \sqrt{m_{\tau}})
\cdot (1,1,1)}{|(\sqrt{m_e} , \sqrt{m_{\mu}} ,
\sqrt{m_{\tau}})||(1,1,1)|}=\frac{\sqrt{m_e}+\sqrt{m_{\mu}}+\sqrt{m_{\tau}}}{\sqrt{3}\sqrt{m_e+m_{\mu}+m_{\tau}}},
\end{equation}
where $\theta_l$ is the angle between the points $(\sqrt{m_e} ,
\sqrt{m_{\mu}} , \sqrt{m_{\tau}})$ and $(1,1,1)$. Commonly, we can
extend $(1,1,1)$ to a more general mass square root vector
$(\sqrt{m_0},\sqrt{m_0},\sqrt{m_0})$, however, the value of $m_0$
does not affect explicitly the value of $\theta_l$. Furthermore, we
can find the relation between $k_l$ and $\theta_l$,
\begin{equation}\nonumber
k_l=\frac{1}{2\cos^2\theta_l},
\end{equation}
and $\theta_l=\frac{\pi}{4}$.

Besides the miraculous success of Koide's relation for charged
leptons, two further questions emerge naturally,

(1). whether this relation can be applied to quarks and neutrinos,

(2). whether this relation holds well at different energy scales.

The first question has been discussed by Esposito and
Santorelli~\cite{Esposito}, Rivero and Gsponer~\cite{Rivero}, and
us~\cite{lima}. In Ref.~\cite{lima}, we applied Koide's relation to
quarks and neutrinos, and estimated the masses of neutrinos. For the
second question, due to the renormalization effect, the masses of
quarks and leptons vary with the energy scale of interaction, so we
must examine Koide's relation at different energy scales explicitly,
before one can take the extension of the relation to quarks and
neutrinos seriously.

In Section II, we check the application of Koide's relation to
quarks and charged leptons. By using the previously introduced
parameters $k_u$, $k_d$ to characterize the deviations of $u$-type
and $d$-type quarks from Koide's relation~\cite{lima}, we find that
$k_u$ and $k_d$ keep almost invariant in a very wide energy range.
This means that Koide's relation is a universal result which is
independent (or insensitive) of the running masses of quarks and
leptons. In Section III, we apply Koide's relation to neutrinos,
with both the normal and inverted mass schemes considered.
Furthermore, with some analogies and conjectures between quarks and
leptons, the neutrino masses are predicted. Finally, in Section IV,
we give some discussions to Koide's relation.

\section{Koide's relation for quarks and charged leptons}

In order to see whether Koide's relation holds well at different
energy scales, we must get the values of quark and lepton masses
first. Because of the renormalization effect, the scale dependence
of the running quark masses is determined by~\cite{Tarasov,koiderun}
\begin{equation}
\mu\frac{d}{d\mu}m(\mu)=-\gamma(\alpha_s)m(\mu),\label{eq.mass}
\end{equation}
and
\begin{eqnarray}
&&\gamma(\alpha_s)=\gamma_0\frac{\alpha_s}{\pi}+\gamma_1\left(\frac{\alpha_s}{\pi}\right)^2
+\gamma_2\left(\frac{\alpha_s}{\pi}\right)^3+O(\alpha_s^4),\nonumber\\
&&\gamma_0=2,\nonumber\\
&&\gamma_1=\frac{101}{12}-\frac{5}{18}n,\nonumber\\
&&\gamma_2=\frac{1}{32}\left[1249-\left(\frac{2216}{27}+\frac{160}{3}\zeta(3)\right)n-\frac{140}{81}n^2\right].\nonumber
\end{eqnarray}
where $\mu$ is an energy scale, $n$ is the effective number of quark
flavors~\cite{Tarasov2}, and $\alpha_s$ is the effective QCD
coupling constant, which is also a $\mu$-dependent function,
\begin{equation}\nonumber
\alpha_s(\mu)=\frac{4\pi}{\beta_0}\frac{1}{L}\left\{1-\frac{2\beta_1}{\beta_0^2}
\frac{\ln L}{L}+\frac{4\beta_1^2}{\beta_0^4L^2}\left[\left(\ln
L-\frac{1}{2}\right)^2
+\frac{\beta_2\beta_0}{8\beta_1^2}-\frac{5}{4}\right]
\right\}+O\left(\frac{\ln^2 L}{L^3}\right),
\end{equation}
where $L=\ln (\mu^2/\Lambda^2)$, and $\beta_0$, $\beta_1$, $\beta_2$
are the coefficients of the renormalization group equation
\begin{equation}
\mu\frac{d}{d\mu}\alpha_s=\beta(\alpha_s),\nonumber
\end{equation}
and
\begin{eqnarray}
&&\beta_0=11-\frac{2}{3}n,\nonumber\\
&&\beta_1=51-\frac{19}{3}n,\nonumber \\
&&\beta_2=2857-\frac{5033}{9}n+\frac{325}{27}n^2.\nonumber
\end{eqnarray}

Using Eq.~(\ref{eq.mass}), the numerical results of the running
masses of quarks were obtained by Fusaoka and
Koide~\cite{koiderun}, as summarized in Table~1, in which both low
and high energy scales are calculated.

First, for the energy scales lower than the spontaneous symmetry
breaking energy scale $\Lambda_W$ of the electroweak gauge symmetry
$SU(2)_L\otimes U(1)_Y$, seven different energy scales are taken
into account, i.e., $\mu=1~\mbox{GeV}$, $\mu=m_c$, $\mu=m_b$,
$\mu=m_W$, $\mu=m_Z$, $\mu=m_t$, and $\mu=\Lambda_W$, where $m_W$
and $m_Z$ are the mass of W and Z bosons.

Second, for the energy scales extremely higher than $\Lambda_W$,
the evolution of the Yukawa coupling constants must be considered.
In~\cite{koiderun}, two different models are presented. One is the
standard model, and the other is the minimal SUSY model. In both
of these two models, the Hamiltonian of the fermion fields can be
written as
\begin{equation}\nonumber
\mathcal{H}=Y^a_{ij}\overline{\psi_{Lai}}\psi_{Raj}\phi^0_a+H.c.,
\end{equation}
where $a=u, d$, $i, j=1, 2, 3$, $Y^a_{ij}$ are the Yukawa coupling
constants, and $\phi^0_a$ are the vacuum expectation values of the
neutral components of the Higgs bosons $\phi_a$. In the standard
model, $\phi^0=\phi^0_u=\phi^0_d$, and in the minimal SUSY model
which has two Higgs bosons, $\phi^0_u=\phi^0\sin \beta(\mu)$ and
$\phi^0_d=\phi^0\cos \beta(\mu)$. The mass matrices at the energy
scale $\mu=\Lambda_W$ of quarks are given by
\begin{equation}\nonumber
m(\mu)_a=\frac{1}{\sqrt{2}}Y(\mu)_av_a,
\end{equation}
where $v_a$ are the vacuum expectation values of $\phi^0_a$, and
$v_a=\sqrt{2}\langle \phi^0_a \rangle$. In the standard model,
$v_u=v_d=\sqrt{2}\Lambda_W$, and in the minimal SUSY model,
$\sqrt{v_u^2+v_d^2}=\sqrt{2}\Lambda_W$.

\begin{table}
\caption {The masses of quarks at different energy scales. The upper
seven rows in Table~I are the cases of lower energy scales, the 8th
and 9th rows are the cases of higher energy scales in the standard
model, and the last two rows are the cases of higher energy scales
in the minimal SUSY model.} \vspace{0.3cm}
\begin{center}
\begin{tabular}{||c||c|c|c||c|c|c||}
  \hline
  \hline
$\mu$ & $m_u~(\mbox{MeV})$ & $m_c~(\mbox{MeV})$ & $m_t~(\mbox{GeV})$ & $m_d~(\mbox{MeV})$ & $m_s~(\mbox{MeV})$ & $m_b~(\mbox{GeV})$\\
  \hline
 $1~\mbox{GeV}$                   & 4.88 & 1506 & 475 & 9.81 & 195.4 & 7.18  \\
 $m_c=1.302~\mbox{GeV}$           & 4.18 & 1302 & 399 & 8.40 & 167.2 & 6.12  \\
 $m_b=4.339~\mbox{GeV}$           & 3.17 & 949  & 272 & 6.37 & 126.8 & 4.34  \\
 $m_W=80.33~\mbox{GeV}$           & 2.35 & 684  & 183 & 4.73 &  94.2 & 3.03  \\
 $m_Z=91.19~\mbox{GeV}$           & 2.33 & 677  & 181 & 4.69 &  93.4 & 3.00  \\
 $m_t=170.8~\mbox{GeV}$           & 2.23 & 646  & 171 & 4.49 &  89.4 & 2.85  \\
 $\Lambda_W=174.1~\mbox{GeV}$     & 2.23 & 645  & 171 & 4.48 &  89.3 & 2.85  \\
  \hline
 $10^9~\mbox{GeV}$              & 1.28 & 371  & 109 & 2.60 &  51.9 & 1.51  \\
 $2\times10^{16}~\mbox{GeV}$    & 0.94 & 272  & 84  & 1.94 &  38.7 & 1.07  \\
  \hline
 $10^9~\mbox{GeV}$              & 1.47 & 427  & 149 & 2.28 &  45.3 & 1.60  \\
 $2\times10^{16}~\mbox{GeV}$    & 1.04 & 302  & 129 & 1.33 &  26.5 & 1.00  \\
  \hline
  \hline
\end{tabular}
\end{center}
\end{table}

We can see from Table~I that all the quark masses decrease with the
increase of the energy scales, and the masses in the minimal SUSY
model are a little bit larger than those in the standard model at
the same energy scales.

Now, we can examine whether Koide's relation holds well for quarks
with the help of the data in Table~I. To characterize the deviation
of quark masses from the exact Koide's relation, here we introduce,
as was done in Ref.~\cite{lima}, two parameters $k_u$ and $k_d$
similarly as in Eq.~(\ref{eq.kl}),
\begin{equation}
k_u\equiv\frac{m_u+m_c+m_t}{\frac{2}{3}(\sqrt{m_u}+\sqrt{m_c}+\sqrt{m_t})^2}
=\frac{1+x_u+y_u}{\frac{2}{3}(1+\sqrt{x_u}+\sqrt{y_u})^2},
\end{equation}
and
\begin{equation}
k_d\equiv\frac{m_d+m_s+m_b}{\frac{2}{3}(\sqrt{m_d}+\sqrt{m_s}+\sqrt{m_b})^2}
=\frac{1+x_d+y_d}{\frac{2}{3}(1+\sqrt{x_d}+\sqrt{y_d})^2},
\end{equation}
where $x_u=m_c/m_u$, $y_u=m_t/m_u$, $x_d=m_s/m_d$, and
$y_d=m_b/m_d$.

With the numerical results of quark masses in Table~I, we can
calculate all the $x_u$, $y_u$, $x_d$ and $y_d$ of both the $u$-type
and $d$-type quarks at different energy scales, and then get $k_u$
and $k_d$ at different energy scales straightforwardly. These
results are listed in Table~II and Table~III.

\begin{table}
\caption{$x_u$, $y_u$ and $k_u$ at different energy scales.}
\vspace{0.3cm}
\begin{center}
\begin{tabular}{||c||c|c|c||c|c|c||}
  \hline
  \hline
    $\mu$ & $m_u~(\mbox{MeV})$ &  $m_c~(\mbox{MeV})$ &  $m_t~(\mbox{GeV})$ &  $x_u$ &  $y_u$ & $k_u$   \\
  \hline
 1~\mbox{GeV}               & 4.88 & 1506 & 475 & 313.75 & 98958 & 1.341  \\
 1.302~\mbox{GeV}           & 4.18 & 1302 & 399 & 311.48 & 95455 & 1.338  \\
 4.339~\mbox{GeV}           & 3.17 & 949  & 272 & 299.37 & 85804 & 1.333  \\
 80.33~\mbox{GeV}           & 2.35 & 684  & 183 & 291.06 & 77872 & 1.328  \\
 91.19~\mbox{GeV}           & 2.33 & 677  & 181 & 290.56 & 77682 & 1.328  \\
 170.8~\mbox{GeV}           & 2.23 & 646  & 171 & 289.69 & 76682 & 1.327  \\
 174.1~\mbox{GeV}           & 2.23 & 645  & 171 & 289.24 & 76682 & 1.327  \\
  \hline
 $10^9~\mbox{GeV}$          & 1.28 & 371  & 109 & 289.84 & 85156 & 1.335  \\
 $2\times10^{16}~\mbox{GeV}$& 0.94 & 272  & 84  & 289.36 & 89362 & 1.339  \\
  \hline
 $10^9~\mbox{GeV}$          & 1.47 & 427  & 149 & 290.48 & 101361 & 1.347  \\
 $2\times10^{16}~\mbox{GeV}$& 1.04 & 302  & 129 & 299.01 & 124038 & 1.359  \\
  \hline
  \hline
\end{tabular}
\end{center}
\end{table}

We can see from Table~II that both $x_u$ and $y_u$ decrease with the
increase of $\mu$ at low energy scales and increase slightly at high
energy scales. However, we find that in spite of the change of $x_u$
and $y_u$, $k_u$ is almost invariant (approximately 1.33) at all
energy scales, which means that $k_u$ is a constant independent of
the running of quark masses. This is an interesting phenomenon, and
indicates that Koide's relation is a universal result.

Furthermore, we can find from Table~II that $k_u$ is not 1 as $k_l$
of charged leptons. This means that Koide's relation should be
improved before being applied to the case of quarks. In~\cite{lima}
we calculated the range of $k_u$, and got $1.1<k_u<1.4$ (with the
mean value o 1.25). However, in~\cite{lima} we did not consider the
renormalization effect of quark masses. With this effect taken into
account, we find that $k_u$ changes a little, from 1.25 to 1.33.

\begin{table}
\caption{$x_d$, $y_d$ and $k_d$ at different energy scales.}
\vspace{0.3cm}
\begin{center}
\begin{tabular}{||c||c|c|c||c|c|c||}
  \hline
  \hline
    $\mu$ & $m_d~(\mbox{MeV})$ & $m_s~(\mbox{MeV})$ & $m_b~(\mbox{GeV})$ & $x_d$ & $y_d$ & $k_d$ \\
  \hline
 1~\mbox{GeV}               & 9.81 & 195.4 & 7.18 & 19.92 & 731.91 & 1.068 \\
 1.302~\mbox{GeV}           & 8.40 & 167.2 & 6.12 & 19.90 & 728.57 & 1.067 \\
 4.339~\mbox{GeV}           & 6.37 & 126.8 & 4.34 & 19.91 & 681.32 & 1.057 \\
 80.33~\mbox{GeV}           & 4.73 &  94.2 & 3.03 & 19.92 & 640.75 & 1.048 \\
 91.19~\mbox{GeV}           & 4.69 &  93.4 & 3.00 & 19.91 & 639.66 & 1.048 \\
 170.8~\mbox{GeV}           & 4.49 &  89.4 & 2.85 & 19.91 & 634.74 & 1.046 \\
 174.1~\mbox{GeV}           & 4.48 &  89.3 & 2.85 & 19.93 & 636.16 & 1.047 \\
  \hline
 $10^9~\mbox{GeV}$          & 2.60 &  51.9 & 1.51 & 19.96 & 580.77 & 1.032 \\
 $2\times10^{16}~\mbox{GeV}$& 1.94 &  38.7 & 1.07 & 19.95 & 551.55 & 1.025 \\
  \hline
 $10^9~\mbox{GeV}$          & 2.28 &  45.3 & 1.60 & 19.87 & 701.75 & 1.062 \\
 $2\times10^{16}~\mbox{GeV}$& 1.33 &  26.5 & 1.00 & 19.92 & 751.88 & 1.072 \\
  \hline
  \hline
\end{tabular}
\end{center}
\end{table}

From Table~III, we can find that $k_d$ also keeps invariant with the
change of energy scales, just as $k_u$, and its approximate value is
1.05.

Similarly, in Table~IV, $k_l$ with the increase of $\mu$ at high
energy scales are listed, both in the standard model and in the
minimal SUSY model, as we know that the values in~\cite{PDG} should
only be taken at low energy scale.

\begin{table}
\caption{$x_l$, $y_l$ and $k_l$ at different energy scales. The
upper three rows are the cases in the standard model, and the
lower three rows are the cases in the minimal SUSY model.}
\vspace{0.3cm}
\begin{center}
\begin{tabular}{||c||c|c|c||c|c|c||}
  \hline
  \hline
   $\mu$ & $m_e~(\mbox{MeV})$ &  $m_{\mu}~(\mbox{MeV})$ &  $m_{\tau}~(\mbox{GeV})$ &  $x_l$ &  $y_l$ & $k_l$   \\
  \hline
 91~\mbox{GeV}              & 0.48684727 & 102.75138  & 1.7467  & 211.05465 & 3587.78 & 1.001881 \\
 $10^9~\mbox{GeV}$          & 0.51541746 & 108.78126  & 1.8492  & 211.05467 & 3587.77 & 1.001881 \\
 $2\times10^{16}~\mbox{GeV}$& 0.49348567 & 104.15246  & 1.7706  & 211.05468 & 3587.95 & 1.001888 \\
  \hline
 91~\mbox{GeV}              & 0.48684727 & 102.75138  & 1.7467  & 211.05465 & 3587.78 & 1.001881 \\
 $10^9~\mbox{GeV}$          & 0.40850306 & 86.21727   & 1.4695  & 211.05661 & 3597.28 & 1.002277 \\
 $2\times10^{16}~\mbox{GeV}$& 0.32502032 & 68.59813   & 1.1714  & 211.05797 & 3604.08 & 1.002560 \\
  \hline
  \hline
\end{tabular}
\end{center}
\end{table}

At the same time, we can see in Table~IV that $k_l$ at high energy
scales is still quite close to 1 as at the low energy scale, which
means that Koide's relation is suitable for charged leptons at all
the energy scales, just as quarks. Also, we can find in Table~IV
that $k_l$ in the minimal SUSY model is a little larger than that in
the standard model.

In summary, we can conclude that Koide's relation is a result
independent (or insensitive) of energy scales of interaction, from
low to extremely high energies.

\section{Koide's relation for neutrinos}

After testing Koide's relation for quarks and charged leptons, and
finding that $k_u$, $k_d$ and $k_l$ are independent of energy
scales, a natural question is what about this relation for
neutrinos. Here we again introduce the parameter $k_{\nu}$ as in
Ref.~\cite{lima}, and discuss this problem. Moreover, we try to find
the relations between the four parameters $k_u$, $k_d$, $k_{\nu}$
and $k_l$, and finally we can get the neutrino masses with some
conjectures.

It is quite difficult to verify whether Koide's relation is
suitable for neutrinos because of the long-existed inaccuracy of
the experimental data of neutrinos. Due to the untiring efforts by
the numerous neutrino experiments, the oscillations and mixings of
neutrinos have been strongly established now. The solar neutrino
deficit is caused by the oscillation from $\nu_{e}$ to a mixture
of $\nu_{\mu}$ and $\nu_{\tau}$ with a mixing angle approximately
of $\theta_{\mathrm{sol}} \approx 34^{\circ}$ in the
KamLAND~\cite{Kam} and SNO~\cite{sno} experiments. Also, the
atmospheric neutrino anomaly is due to the $\nu_{\mu}$ to
$\nu_{\tau}$ oscillation with almost the largest mixing angle of
$\theta_{\mathrm{atm}} \approx 45^{\circ}$ in the K2K~\cite{K2K}
and Super-Kamiokande~\cite{SUPER} experiments. However, the
non-observation of the disappearance of $\bar{\nu}_{e}$ in the
CHOOZ~\cite{Chz} experiment showed that the mixing angle
$\theta_{\mathrm{chz}}$ is smaller than $5^{\circ}$ at the best
fit point~\cite{garcia,Altarelli}.

These experiments also measured the mass-squared differences of
the neutrino mass eigenstates. According to the global analysis of
the experimental results, we have (the allowed ranges at
$3\sigma$)~\cite{Altarelli}
\begin{equation}
1.4\times10^{-3}~\mbox{eV}^{2}<\Delta
m_{\mathrm{atm}}^{2}=|m_{3}^{2}-m_{2}^{2}|<3.7\times10^{-3}~\mbox{eV}^{2},\label{eq.3}
\end{equation}
and
\begin{equation}
5.4\times10^{-5}~\mbox{eV}^{2}<\Delta
m_{\mathrm{sol}}^{2}=|m_{2}^{2}-m_{1}^{2}|<9.5\times10^{-5}~\mbox{eV}^{2},\label{eq.4}
\end{equation}
where $m_1$, $m_2$, $m_3$ are the masses of the three mass
eigenstates of neutrinos, and the best fit points are
$|m_{3}^{2}-m_{2}^{2}|=2.6\times 10^{-3}~\mbox{eV}^{2}$, and
$|m_{2}^{2}-m_{1}^{2}|=6.9\times 10^{-5}~\mbox{eV}^{2}$
\cite{Altarelli}.

Due to Mikheyev-Smirnov-Wolfenstein~\cite{msw} matter effect of
solar neutrinos, we already know that $m_2>m_1$. Hence we have
\begin{equation}
m_{1}^{2}=m_{2}^{2}-\Delta m_{\mathrm{sol}}^{2},\label{eq.5}
\end{equation}
and
\begin{equation}
m_{3}^{2}=m_{2}^{2}\pm\Delta m_{\mathrm{atm}}^{2}.\label{eq.6}
\end{equation}
So there are two mass schemes, (1) the normal mass scheme
$m_3>m_2>m_1$, and (2) the inverted mass scheme $m_2>m_1>m_3$. We
will discuss both of them in the following.

Now we apply Koide's relation to neutrinos. First we take the
normal mass scheme for example. If Koide's relation holds well for
neutrinos, we have
\begin{equation}
m_1+m_2+m_3=\frac{2}{3}\left(\sqrt{m_1}+\sqrt{m_2}+\sqrt{m_3}\right)^2.\label{eq.nu}
\end{equation}
Substituting Eqs.~(\ref{eq.5}) and~(\ref{eq.6}) into
Eq.~(\ref{eq.nu}), we get
\begin{equation}\nonumber
\sqrt{m^2_2-\Delta m_{\mathrm{sol}}^{2}}+m_2+\sqrt{m^2_2+\Delta
m_{\mathrm{atm}}^{2}}=\frac{2}{3}\left(\sqrt[4]{m^2_2-\Delta
m_{\mathrm{sol}}^{2}}+\sqrt{m_2}+\sqrt[4]{m^2_2+\Delta
m_{\mathrm{atm}}^{2}}\right)^2.
\end{equation}

Solving this equation, we find that there is no real root for
$m_2$ with the restrictions in Eqs.~(\ref{eq.3}) and~(\ref{eq.4}).
This means that no matter what value $m_2$ is, Koide's relation
does not hold for neutrinos. So is the inverted mass scheme.

Thus we must introduce a parameter $k_{\nu}$~\cite{lima} to
character the deviation of neutrinos from Koide's relation,
\begin{equation}
k_{\nu}\equiv\frac{m_1+m_2+m_3}{\frac{2}{3}(\sqrt{m_1}+\sqrt{m_2}+\sqrt{m_3})^2}.
\end{equation}

Since $k_{\nu}\not=1$, we must determine its range, and this can
help us to find the relations between $k_u$, $k_d$, $k_{\nu}$ and
$k_l$, and fix the neutrino masses. We now check the situations
for the two mass schemes, respectively.

1. For the normal mass scheme, $m_3>m_2>m_1$, we have
\begin{equation}
k_{\nu}=\frac{\sqrt{m^2_2-\Delta
m_{\mathrm{sol}}^{2}}+m_2+\sqrt{m^2_2+\Delta
m_{\mathrm{atm}}^{2}}}{\frac{2}{3}\left(\sqrt[4]{m^2_2-\Delta
m_{\mathrm{sol}}^{2}}+\sqrt{m_2}+\sqrt[4]{m^2_2+\Delta
m_{\mathrm{atm}}^{2}}\right)^2}.\label{eq.10}
\end{equation}
We can see that $k_{\nu}$ is the function of $m_2$ if $\Delta
m_{\mathrm{sol}}^{2}$ and $\Delta m_{\mathrm{atm}}^{2}$ are fixed.
Due to the inaccuracy of the experimental data, we take $\Delta
m_{\mathrm{sol}}^{2}$ and $\Delta m_{\mathrm{atm}}^{2}$ as their
best fit values here. The range of $k_{\nu}$ is shown in Fig.~1.

\begin{figure}
\begin{center}
\scalebox{0.45}{\includegraphics{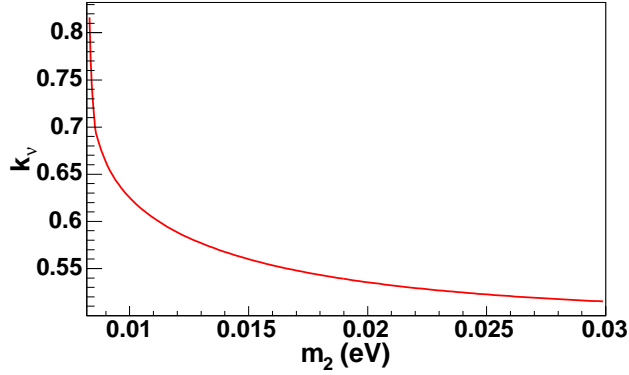}}
\end{center}
\caption{The range of $k_{\nu}$ of the normal mass scheme
$m_3>m_2>m_1$.}\label{fig1}
\end{figure}

We can see from Fig.~1 that $0.50<k_{\nu}<0.85$, and $k_{\nu}$
decreases with the increase of $m_2$. So $k_{\nu}<1$ for
neutrinos.

2. For the inverted mass scheme, $m_2>m_1>m_3$, we have
\begin{equation}
k_{\nu}=\frac{\sqrt{m^2_2-\Delta
m_{\mathrm{sol}}^{2}}+m_2+\sqrt{m^2_2-\Delta
m_{\mathrm{atm}}^{2}}}{\frac{2}{3}\left(\sqrt[4]{m^2_2-\Delta
m_{\mathrm{sol}}^{2}}+\sqrt{m_2}+\sqrt[4]{m^2_2-\Delta
m_{\mathrm{atm}}^{2}}\right)^2}.
\end{equation}
The range of $k_{\nu}$ is shown in Fig.~2.

\begin{figure}
\begin{center}
\scalebox{0.45}{\includegraphics{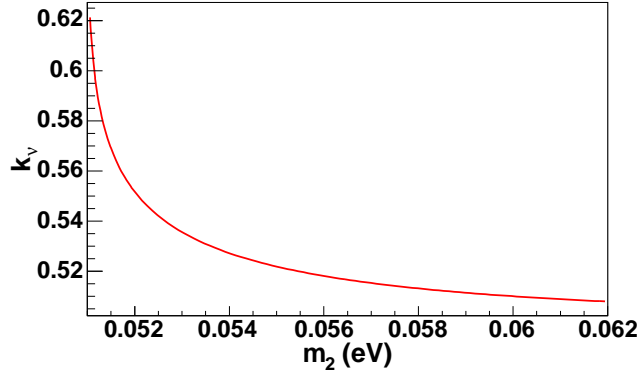}}
\end{center}
\caption{The range of $k_{\nu}$ of the inverted mass scheme
$m_2>m_1>m_3$.}\label{fig2}
\end{figure}

We can see from Fig.~2 that $0.50<k_{\nu}<0.65$. Again, $k_{\nu}<1$
for neutrinos.

Altogether, $0.50<k_{\nu}<0.85$ for both of these two mass
schemes. And $k_{\nu}$ of the normal mass scheme is larger than
that of the inverted mass scheme.

Conclusively, the values of $k_u$, $k_d$, $k_{\nu}$ and $k_l$ can be
summarized as follows,
\begin{equation}
 \left(
    \begin {array}{c}
       \nu_{e} \\
       e
\end{array}
\right) \left(
    \begin {array}{c}
       \nu_{\mu} \\
       \mu
\end{array}
\right) \left(
    \begin {array}{c}
       \nu_{\tau} \\
       \tau
\end{array}
\right)
\begin {array}{c}
k_{\nu}<1\\
k_{l}=1
\end{array},\quad \mbox{and} \quad
\left(
    \begin {array}{c}
       u \\
       d
\end{array}
\right) \left(
    \begin {array}{c}
       c \\
       s
\end{array}
\right) \left(
    \begin {array}{c}
       t \\
       b
\end{array}
\right)
\begin {array}{c}
k_{u}>1\\
k_{d}\approx1
\end{array}.
\end{equation}

We believe that the problem of the origin of the masses of leptons
must be solved together with that of quarks. Since $k_{l}=1$ and
$k_{d}\approx1$, we may conjecture that $k_{l}+k_{d}\approx 2$. At
the same time, since $0.50<k_{\nu}<0.85$ and $k_{u}\approx1.33$,
we may analogize the conjecture of $k_{l}$ and $k_{d}$, and
propose the hypothesis that
\begin{equation}
k_{\nu}+k_{u}\approx 2.\label{eq.kuknu}
\end{equation}
This is from the speculation that there must be some relation
between $k_u$, $k_d$, $k_{\nu}$ and $k_l$. Of course, this Ansatz
is not the unique one of the relations between $k_u$, $k_d$,
$k_{\nu}$ and $k_l$. For example, we may also assume that
\begin{equation}
k_l k_d \approx k_{\nu}k_u \approx 1,\label{eq.kk}
\end{equation}
or
\begin{equation}
\frac{1}{k_l}+\frac{1}{k_d} \approx
\frac{1}{k_{\nu}}+\frac{1}{k_u}\approx 2. \label{eq.1k}
\end{equation}
Eq.~(\ref{eq.1k}) is based on the assumption that $\theta_l+\theta_d
\approx \theta_{\nu}+\theta_u \approx \frac{\pi}{2}$ in Foot's
geometrical interpretation.

All these hypotheses can show the balance between $k_{\nu}$ and
$k_u$ intuitively and transparently. This situation seems to be
similar to the quark-lepton complementarity between mixing angles
of quarks and leptons~\cite{Complementarity}, and we may call it a
quark-lepton complementarity of masses.


From Table~II, we can see that the value of $k_{u}$ is approximately
1.33. Thus from the hypothesis $k_{\nu}+k_{u}\approx2$, we get that
$k_{\nu}\approx 0.67$. This is consistent with the normal mass
scheme and in conflict with the inverted mass scheme. This indicates
that the three masses of neutrino mass eigenstates are heavier in
order, which is the same as quarks and charged leptons.

Now we can estimate the absolute masses of neutrinos. Substituting
$k_{\nu}=0.67$, $\Delta m_{\mathrm{atm}}^{2}=2.6\times
10^{-3}~\mbox{eV}^{2}$, and $\Delta m_{\mathrm{sol}}^{2}=6.9\times
10^{-5}~\mbox{eV}^{2}$ into Eq.~(\ref{eq.10}), we obtain the value
of $m_2$,
\begin{equation}
0.67=\frac{\sqrt{m^2_2-6.9\times
10^{-5}~\mbox{eV}^{2}}+m_2+\sqrt{m^2_2+2.6\times
10^{-3}~\mbox{eV}^{2}}}{\frac{2}{3}\left(\sqrt[4]{m^2_2-6.9\times
10^{-5}~\mbox{eV}^{2}}+\sqrt{m_2}+\sqrt[4]{m^2_2+2.6\times
10^{-3}~\mbox{eV}^{2}}\right)^2},\label{eq.18}
\end{equation}
and we get $m_2=0.0089~\mbox{eV}$.

Straightforwardly, we get
\begin{equation}
m_1=\sqrt{m^2_2-\Delta
m_{\mathrm{sol}}^{2}}=0.0032~\mbox{eV},\label{eq.19}
\end{equation}
and
\begin{equation}
m_3=\sqrt{m^2_2+\Delta
m_{\mathrm{atm}}^{2}}=0.052~\mbox{eV}.\label{eq.20}
\end{equation}


Now we discuss the uncertainty of $m_1$, $m_2$ and $m_3$. In
Fig.~1, we can see that the slope of the curve in very large where
$0.65<k_{\nu}<0.8$, so the value of $m_2$ is not very sensitive to
the error of $k_{\nu}$. $m_2$ will approximately be $8.5\sim
8.9\times10^{-3}~\mbox{eV}$ even if the value of $k_{\nu}$ charges
from 0.65 to 0.8, so the value of $m_2$ is precise to a fairly
good degree of accuracy. Similarly, the value of $m_3$ will be
about $0.052~\mbox{eV}$ to a good degree of accuracy too, because
$m_3=\sqrt{m^2_2+\Delta m_{\mathrm{atm}}^{2}}$, and $\Delta
m_{\mathrm{atm}}^{2} \gg m^2_2$. The only point desired to be
mentioned here is the range of $m_1$. Because if $m^2_2$ is rather
closed to $\Delta m_{\mathrm{sol}}^{2}$, and due to the big
uncertainty of $\Delta m_{\mathrm{sol}}^{2}$, the value of $m_1$
may change largely with $k_{\nu}$. We can see this in the other
two hypotheses in Eqs.~(\ref{eq.kk}) and (\ref{eq.1k}).

1. In Eq.~(\ref{eq.kk}), where $k_l k_d \approx k_{\nu}k_u \approx
1$, we have $k_{\nu}\approx 0.75$, and
\begin{eqnarray}
&&m_1=0.0012~\mbox{eV},\nonumber\\
&&m_2=0.0084~\mbox{eV},\nonumber\\
&&m_3=0.050~\mbox{eV}.\label{eq.kkm}
\end{eqnarray}

2. In Eq.~(\ref{eq.1k}), where $\frac{1}{k_l}+\frac{1}{k_d}
\approx \frac{1}{k_{\nu}}+\frac{1}{k_u}\approx 2$, we have
$k_{\nu}\approx 0.80$, and
\begin{eqnarray}
&&m_1=1.0\times10^{-5}~\mbox{eV},\nonumber\\
&&m_2=0.0084~\mbox{eV},\nonumber\\
&&m_3=0.050~\mbox{eV}.\label{eq.1km}
\end{eqnarray}

From Eqs.~(\ref{eq.19}), (\ref{eq.kkm}) and (\ref{eq.1km}), we can
see that the value $m_1$ is only a rough estimate of the first
step till now, and its effective number and order of magnitude may
change with the more and more precise experimental data in the
future. However, the values of $m_2$ and $m_3$ are consistent in
these three hypotheses.

Finally, we should point out that Koide~\cite{koide3} also gave an
interpretation of his relation as a mixing between octet and singlet
components in a nonet scheme of the flavor $U(3)$, and got the
neutrino masses as $m_1=0.0026~\mbox{eV}$, $m_2=0.0075~\mbox{eV}$
and $m_3=0.050~\mbox{eV}$~\cite{koide4}. We can see that his results
are strongly consistent with ours, especially with the results from
Eq.~(\ref{eq.kuknu}). And recently, he got the three masses as
$m_1=0.0039~\mbox{eV}$, $m_2=0.0088~\mbox{eV}$ and
$m_3=0.053~\mbox{eV}$ in a seesaw mass matrix model of quarks and
leptons with flavor-triplet Higgs scalars~\cite{Koideseesaw}, which
is even closer to our results from Eq.~(\ref{eq.kuknu}).

\section{Summary}

Finally, we give some discussions.

1. Since Koide's relation is such a wonderful result for charged
leptons at low energy scale, to explore its behavior at high energy
is straightforward. We carefully test whether it is energy scale
independent in Section II, and find that it is really independent or
insensitive of energy scale in a huge energy range and is almost the
same in both the standard model and the minimal SUSY model, which
proves that Koide's relation is a universal result in particle
physics.

2. Seesaw mechanism~\cite{seesaw} may give the origin of neutrino
masses, in which the right-handed very-heavy neutrinos is included
into the Lagrangian of the standard model, i.e.,

\begin{eqnarray}
\left(
      \begin{array}{cc}
      \overline{\nu_L} & \overline{\nu_L}^C  \\
      \end{array}
        \right)
\left(
      \begin{array}{cc}
      0 & m_D\\
      m_D^T & M_R\\
      \end{array}
        \right)
\left(
\begin{array}{cc}
      \nu_R^C  \\
      \nu_R \\
       \end{array}
        \right),\label{eq.seesaw}
\end{eqnarray}
where the scale of $m_D$ is characterized to be the energy scale $v$
of the electroweak spontaneous breaking, and $M_R$ is the
right-handed very-heavy neutrino mass matrix.

From Eq.~(\ref{eq.seesaw}), we can get
\begin{equation}\nonumber
m_{\nu}=-m_DM_Rm_D^T,
\end{equation}
where the neutrino mass eigenvalues are $m_1=m_u^2/M_1$,
$m_2=m_c^2/M_2$ and $m_3=m_u^2/M_3$. Thus, the smallness of neutrino
masses is due to the large values of $M_1$, $M_2$ and $M_3$. (For
example, $m_{\nu}\sim 0.1~\mbox{eV}$ if $M_R$ is taken as
$10^{14}~\mbox{GeV}$.)

However, seesaw mechanism can only present an illustrative
interpretation of the origin of neutrino masses, without accurate
prediction of the masses of neutrino mass eigenstates. To obtain
those, we must extend our theory and make some speculation, i.e., we
examine whether Koide's relation of charged leptons also holds well
for neutrinos, and we find that not all quarks and leptons obey
Koide's relation precisely. So we introduce the parameters $k_u$,
$k_d$ and $k_{\nu}$ to describe the deviations of quarks and
neutrinos from the exact Koide's relation. With this improved
relation and the conjecture of a quark-lepton complementarity of
masses such as $k_{l}+k_{d}\approx k_{\nu}+k_{u}\approx2$, $k_l k_d
\approx k_{\nu}k_u \approx 1$ or $\frac{1}{k_l}+\frac{1}{k_d}
\approx \frac{1}{k_{\nu}}+\frac{1}{k_u}\approx 2$, we can determine
the absolute masses of the neutrino mass eigenstates. Due to the
inaccuracy of the experimental data of neutrinos nowadays, these
results (especially the value of $m_1$) should be only taken as
primary estimates. It is also possible that seesaw mechanism is
responsible for the deviation of the neutrino masses from the exact
Koide's relation.

3. There remain some open questions to be answered. Such as

(1). Which hypothesis between $k_u$, $k_d$, $k_{\nu}$ and $k_l$ is
really the relation between them? For example, if
$\frac{1}{k_l}+\frac{1}{k_d} \approx
\frac{1}{k_{\nu}}+\frac{1}{k_u}\approx 2$, we have
$\theta_l+\theta_d \approx \theta_{\nu}+\theta_u \approx
\frac{\pi}{2}$ in Foot's geometrical interpretation, is there
really some deeper reason behind it, just like the quark-lepton
complementarity of their mixing angles~\cite{Complementarity}?

(2). We can see from Table~II that $k_u$ is approximately 1.33, so
may it be $\frac{4}{3}$ exactly? If so, we have
$k_{\nu}=\frac{2}{3}$ in Eq.~(\ref{eq.kuknu}), $k_{\nu}=\frac{3}{4}$
in Eq.~(\ref{eq.kk}), and $k_{\nu}=\frac{4}{5}$ in
Eq.~(\ref{eq.1k}). All these $k_u$ and $k_{\nu}$ are of special
values, and is there some unknown principle leading this? Moreover,
is $k_d=1$ exactly as $k_l$, or deviates from 1 slightly? If so,
why?

In conclusion, we believe that there must be some deeper principle
behind the elegant Koide's relation for charged leptons, and thus it
is meaningful to check whether this relation is also applicable to
quarks and neutrinos, at least at some degree. We show in this paper
that the Koide's relation with its deviation characterized by the
parameters $k_u$ and $k_d$ is also applicable to quarks without
sensitivity to energy scale. By using the improved Koide's relation
with an Ansatz of quark-lepton complementarity of masses, we can
also determine neutrino masses. If the prediction of neutrino masses
is tested to be consistent with the precise experiments in the
future, it would be a big success of Koide's relation, which may
shed light on our way to a grand unification theory of quarks and
leptons.

{\bf Acknowledgments}

We are very grateful to Profs.~Xiao-Gang He, Yoshio Koide, Zhi-Zhong
Xing and Daxin Zhang for their stimulating suggestions and valuable
discussions. We also thank Xun Chen and Xiaorui Lu for discussions.
This work is partially supported by National Natural Science
Foundation of China (Nos.~10421503, 10575003, 10505001, 10528510),
by the Key Grant Project of Chinese Ministry of Education
(No.~305001), by the Research Fund for the Doctoral Program of
Higher Education (China).



\begin{thebibliography}{99}
\bibitem{barut}
A.O.~Barut, \Journal{\PRL} {42}{1251}{1979}.

\bibitem{hdv}
http://www.physcomments.org/wiki/index.php?title=Bakery:HdV.

\bibitem{koide}
Y.~Koide, \Journal{\LNC} {34}{201}{1982};

Y.~Koide, \Journal{\PLB} {120}{161}{1983}.

\bibitem{koide2}
Y.~Koide, \Journal{\PRD} {28}{252}{1983}.

\bibitem{koideother}
Y.~Koide, \Journal{\ZPC} {45}{39}{1989};

Y.~Koide, \Journal{\MPLA} {5}{2319}{1990};

Y.~Koide, H.~Fusaoka, \Journal{\ZPC} {71}{459}{1996};

Y.~Koide, M.~Tamimoto, \Journal{\ZPC} {72}{333}{1996};

Y.~Koide, \Journal{\PRD} {60}{077301}{1999}.

\bibitem{review}
Y.~Koide, hep-ph/0506247.

\bibitem{PDG}
Particle Data Group, K.~Hagiwara, {\it et al.}, \Journal{\PRD}
{66}{010001}{2002}.

\bibitem{foot}
R.~Foot, hep-ph/9402242.

\bibitem{Esposito}
S.~Esposito, P.~Santorelli, \Journal{\MPLA} {10}{3077}{1995}.

\bibitem{Rivero}
A.~Rivero, A.~Gsponer, hep-ph/0505220.

\bibitem{lima}
N.~Li, B.-Q.~Ma, \Journal{\PLB} {609}{309}{2005}.

\bibitem{Tarasov}
O.V.~Tarasov, Dubna Report No.~JINR P2-82-900, 1982 (unpublished).

\bibitem{koiderun}
H.~Fusaoka, Y.~Koide, \Journal{\PRD} {57}{3986}{1998}.

\bibitem{Tarasov2}
O.V.~Tarasov, A.A.~Vladimirov, A.Yu.~Smirnov, \Journal{\PLB}
{93}{429}{1980}.

\bibitem{Kam}
KamLAND Collaboration, K.~Eguchi, {\it et al.},  \Journal{\PRL}
{90}{021802}{2003}.

\bibitem{sno}
SNO Collaboration, S.N.~Ahmad, {\it et al.}, \Journal{\PRL}
{92}{181301}{2004}.

\bibitem{K2K}
K2K Collaboration, M.H.~Ahn, {\it et al.},  \Journal{\PRL}
{90}{041801}{2003}.

\bibitem{SUPER}
Super-Kamiokande Collaboration, Y.~Fukuda, {\it et al.}, Phys.
Rev. Lett. {\bf 81}, 1562 (1998);  Y.~Ashie, {\it et al.}, Phys.
Rev. Lett. {\bf 93}, 101801 (2004).

C.K.~Jung, C.~McGrew, T.~Kajita, T.~Mann, Anna. Rev. Nucl. Part.
Sci. {\bf 51}, 451 (2001).

\bibitem{Chz}
CHOOZ Collaboration, M.~Apollonio, {\it et al.},
\Journal{\PLB}{420}{397}{1998};

Palo Verde Collaboration, F.~Boehm, {\it et al.}, \Journal{\PRL}
{84}{3764}{2000}.

\bibitem{garcia}
M.C.~Gonzalez-Garcia, hep-ph/0410030.

\bibitem{Altarelli}
G.~Altarelli, hep-ph/0410101.

\bibitem{msw}
S.P.~Mikheyev, A.Yu.~Smirnov, \Journal{Sov. J. Nucl. Phys.}
{42}{913}{1985};

L.~Wolfenstein, \Journal{\PRD}{17}{2369}{1978}.




\bibitem{Complementarity}
A.Yu.~Smirnov, hep-ph/0402264;

M.~Raidal, Phys. Rev. Lett. {\bf 93}, 161801 (2004);

H.~Minakata, A.Yu.~Smirnov, Phys. Rev. {\bf D 70}, 073009 (2004);

N.~Li, B.-Q.~Ma, \Journal{\PRD}{71}{097301}{2005};

N.~Li, B.-Q.~Ma, \Journal{\EPJC}{42}{17}{2005};

T.~Ohlsson,
\Journal{\PLB}{622}{159}{2005}.

\bibitem{koide3}
Y.~Koide, M.~Tanimoto, \Journal{Z. Phys. C} {72}{333}{1996}.

\bibitem{koide4}
Y.~Koide, H.~Nishiura, K.~Matsuda, T.~Kikuchi, T.~Fukuyama,
\Journal{\PRD} {66}{093006}{2002}.

\bibitem{Koideseesaw}
Y.~Koide, hep-ph/0508301.


\bibitem{seesaw}
T.~Yanagida, in {\it Proceedings of the Workshop on Unified Theory
and the Baryon Number of the Universe}, edited by O.~Sawada and
A.~Sugamoto (KEK, Tsukuba, 1979);

M.~Gell-Mann, P.~Ramond, R.~Slansky, in {\it Supergravity}, edited
by F.~van Nieuwenhuizen and D.~Freedman (North Holland, Armsterdam,
1979);

R.N.~Mohapatra, G.~Senjanovic, Phys. Rev. Lett. {\bf 44}, 912
(1980).



\end{thebibliography}
\end{document}